# Controlled inter-state switching between quantized conductance states in resistive devices for multilevel memory

Sweety Deswal,[†,‡] Rupali R. Malode,[§] Ashok Kumar,[†,‡] and Ajeet Kumar*[,†,‡]

[†]Academy of Scientific and Innovative Research (AcSIR), Ghaziabad- 201002, India.

[‡]CSIR-National Physical Laboratory, Dr. K.S. Krishnan Marg, New Delhi 110012, India.

[§]Maulana Azad National Institute of Technology, Bhopal, Madhya Pradesh 462003, India.

*Email: kumarajeet@nplindia.org

**ABSTRACT**: A detailed understanding of quantization conductance (QC), their correlation with resistive switching phenomena and controlled manipulation of quantized states is crucial for realizing atomic-scale multilevel memory elements. Here, we demonstrate highly stable and reproducible quantized conductance states (QC-states) in Al/Niobium oxide/Pt resistive switching devices. Three levels of control over the QC-states, required for multilevel quantized state memories, like, switching ON to different quantized states, switching OFF from quantized states, and controlled inter-state switching among one QC-states to another has been demonstrated by imposing limiting conditions of stop-voltage and current compliance. The well defined multiple QC-states along with a working principle for switching among various states show promise for implementation of multilevel memory devices.

Driven by the demand for improved computing capability, the semiconductor industry is following the extension of Moore's law which says that the density of transistors in an integrated circuit doubles every two years. However the current technology, charge based flash memory, has reached its limits of miniaturization.[1-2] Also, all existing devices are limited to two stable memory states (i.e., "0" and "1"). Increasing the number of stable states, from bi-stability to multi-stability, will be an effective method for producing high-density and efficient memory devices.

As one for the most promising candidate for future non-volatile memories, resistive random access memory (ReRAM) with simple two-terminal sandwiched structured devices exhibit attractive performances due to their scalability down to atomic level, CMOS compatibility, low-power consumption, and high-speed features.[3-4] It has been proposed that the multiple stable states available in the resistive switches can be used for multilevel storage for ultrahigh density memories.[5] Existence of stable multistates has been demonstrated in resistive switching,[6-14] ferroelectric[15-20] and phase change[21-25] memory devices. Atomic point contact based QC observed in resistive switching devices has also been demonstrated for memory applications.[26-31] However, controlled manipulation of multiple stable states for potential application in multilevel memory is yet to be achieved.

Several kinds of control over stable QC-states in a resistive switching device are required to achieve multilevel quantized state memories. These particular kinds of devices have not been fully explored, partly because of the lack of appropriate materials and lack of design & working principles. Many groups have demonstrated quantization in several ReRAM[32-36] as well as in atomic switch[29,37-39] devices. The conditions to achieve different quantized states either with current compliance[40-41] or with stop voltage[37,40] have been reported. Also, there is some understanding about the stability of these states with respect to time.[35,37,42-43] However, conditions for controlled inter-QC-state switching, essential for multilevel memory, have not been reported.

Here, we demonstrate control over the events of switching ON to different QC-states, switching OFF from QC-states, and inter-QC-state switching in Al/Niobium oxide/Pt device. Firstly, stable and reproducible QC-states with integer and half-integer multiples of quantum of conductance ($G_0 = 2e^2/h \sim 77.4$ μS) were achieved, indicating formation of well-controlled atomic point contacts in the conducting filaments. Then, the devices were manipulated to exhibit hundreds of different inter-QC-state switching, both in the direction of SET (higher $G_0$) or RESET (lower $G_0$) starting from any particular QC-state. The initial and final QC-states, for each switching event, were found to be stable. The device exhibited longer retention times for higher QC-states. Rules for controlled switching are evolved with stop-voltage and current compliance limits during current-voltage (*I-V*) measurements. The working principles demonstrated in this work, to achieve QC-states and to induce inter-QC-state switching, is a crucial step towards realization of multilevel memory devices.

*Switching ON to QC-state*: The resistive switching and QC characteristics are demonstrated using *I-V* measurements on Al/Nb$_2$O$_5$/Pt devices in air at room temperature. These devices, in their pristine state, were found in high resistance OFF state (HRS) of the order of $\sim 10^9$ Ω. Initially, the device was switched to low resistance ON state (LRS) at a forming voltage ~4 V with current compliance ($I_c$) of 5 μA, as shown in the inset of Figure 1a. After forming, with voltage sweeps, the device showed reproducible switching between LRS to HRS (RESET; voltage ~ -0.4 to -1.2 V) and vice-versa (SET; voltage ~1.6-2.5 V), shown as semi-logarithmic *I-V* plots in Figure 1a. These devices show both unipolar as well as bipolar switching characteristics in either polarities of the voltage (Figure S1). In our previous work,[34] unipolar switching behaviour of the Al/Nb$_2$O$_5$/Pt devices were presented and it was demonstrated that the conducting filament, after making the atomic point contact, grows in thickness atom-by-atom during SET voltage sweep. Here, in this work, conducting filaments were stabilized to achieve various QC-states. During the SET process, the LRS was controlled by applying voltage sweeps with different current compliance values of 100, 200, 300, 400 and 500 µA (Figure 1b) and different resistance states of 9 kΩ, 6 kΩ, 4 kΩ, 2.9 kΩ and 2.3 kΩ, respectively, were achieved. These resistance states were stable and correspond to quantized conductance states of ~1.5 $G_0$, ~2 $G_0$, ~3.5 $G_0$, ~4.5 $G_0$, and ~5.5 $G_0$, respectively (Figure 1c).

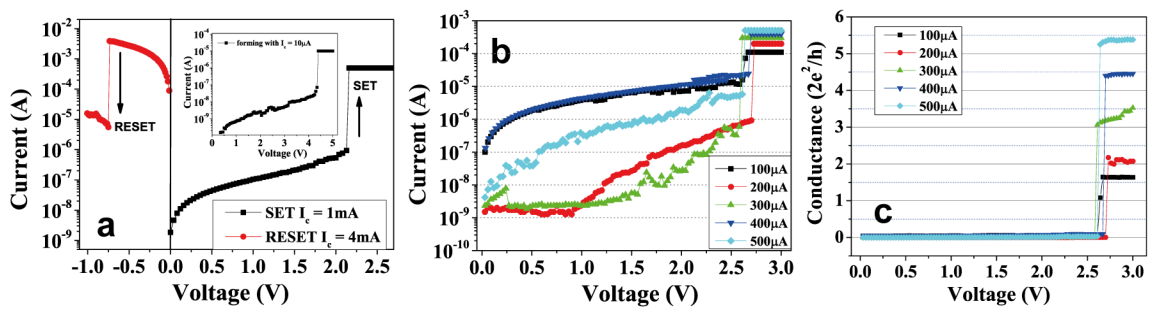

**Figure 1.** (a) Semi-logarithmic *I-V* characteristics of Al/Nb$_2$O$_5$/Pt device showing bipolar switching with SET and RESET in range of 1.6-2.5 V and –(0.4-1.2 V), respectively. The inset shows the electroformation step of the device. (b) The semi-logarithmic *I-V* plots of SET with various current compliances ($I_c$) values of 100, 200, 300, 400, 500 µA reaching to different LRS levels corresponding to quantized conductance states of ~1.5 $G_0$, ~2 $G_0$, ~3.5 $G_0$, ~4.5 $G_0$, and ~5.5 $G_0$, respectively. (c) The SET traces of (b) are plotted as conductance vs. voltage (*G-V*) to show distinguishable quantized LRS states obtained.

QC-states achieved during SET sweeps with different $I_c$ values were analyzed to determine the state distribution of the device conductance. Figures 2a-f show that distinct and stable QC- states could be reproducibly achieved by varying $I_c$ values. Histograms of conductance in the units of $G_0$ for ~300 switching cycles performed on an Nb$_2$O$_5$ device are shown for five different compliance currents upto 500 µA. After each SET event, the conductance state was estimated by applying a read voltage of 100 mV. The data was sorted in the bin size of 0.1 $G_0$ and respective numbers were counted to plot the conductance histogram, shown in Figure 2. QC-state of ~1 $G_0$ was achieved with 100 µA, where only a single conduction channel allows electron transport through the filament of the resistive switch. With the increase of current compliance, the conductance peak shifted towards higher conductance value. The QC-states of 2 $G_0$, 3.5 $G_0$, 4.5 $G_0$, and 5.5 $G_0$ were achieved with $I_c$ of 200 μA, 300 μA, 400 μA, and 500 μA, respectively, as seen in Figure 2b-e. As higher conductance states are gradually reached, it has been understood that the atomic rearrangements in the point contact allows more number of conduction channels to become available for electron transfer.[31,34] The histogram with all five sets of each $I_c$, acquired from ~300 curves of SET cycles is shown in Figure 2f. It can be clearly seen that the devices exhibited quantized conductance peaks around integer and half-integer multiples of $G_0$. Out of several switching cycles, 25 cycles with each $I_c$ values 100, 200, 300, 400, 500 μA representing the median of the distribution of quantized state as plotted in Figures 2a-e is shown in Figure 2g. The well separated memory levels available in our devices meet one of the essential requirements for realizing multilevel ultra high density storage.

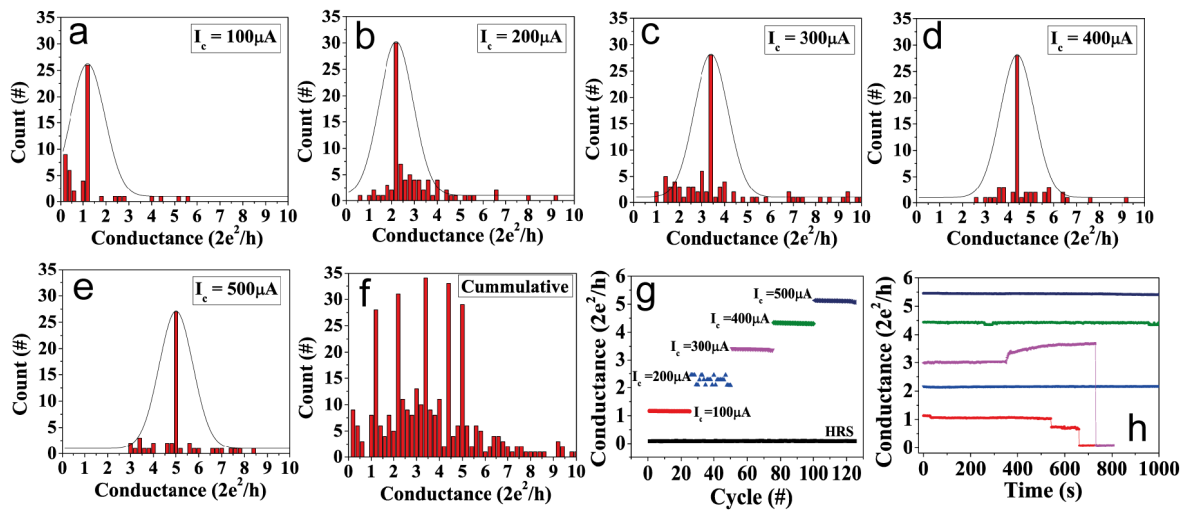

**Figure 2.** The histogram of quantized conductance values obtained during SET with $I_c$ of (a) 100 μA (b) 200 μA, (c) 300 μA, (d) 400 μA, (e) 500 μA. Each plot shows the data of more than 60 cycles of SET for a particular $I_c$. (f) The cumulative data of all $G_0$ (in Figure 2a-e) obtained for ~300 cycles of SET. (g) The median of the distribution shows distinct conductance states after SET with current compliance of 100, 200, 300, 400, 500 μA along with HRS is exhibited for 125 switching cycles. (h) The stability of corresponding quantized conductance states shown in Figure (g).

*Stability of QC-states & retention time*: To understand the stability of the quantized conductance states, retention time characteristics of different conductance states were studied. Different QC-states were achieved in different SET sweeps and their retention time was measured at 100 mV read voltage. Figure 2h shows the retention time of >500 s for QC-states corresponding to 1 $G_0$, 2 $G_0$, 3.5 $G_0$, 4.5 $G_0$, and 5.5 $G_0$. The retention time of different QC-states were observed to be increasing with increase in $G_0$. In general, QC-states below 3 $G_0$ were stable for less than 800 s, while the QC-states higher than 3 $G_0$ were stable for more than 1000 s. However, on some occasions, stability over 1000 s were also observed for states <3 $G_0$. Retention data of various other QC-states are shown in supplementary Figure S2a. The stability of a particular QC-state depends on the strength of the corresponding conducting filament. The conducting filament diameter increases as the $G_0$ of QC-states increase, thus making them more and more robust. The magnitude of applied read voltage during retention measurement was also found to influence the stability of QC-states (supplementary fig2b).

*Inter-QC-state switching*: Once a device is switched ON to a particular QC-state, voltage sweep and current compliance conditions could be controlled to exhibit many different inter-QC-state switching in the device, be in the direction of SET (higher $G_0$) or RESET (lower $G_0$). Figure 3 shows one set of four successive switching steps of inter-QC-state in SET direction of a particular device along with the corresponding QC-state retention time up to 100 s. The device was, firstly, SET to ~2.5 $G_0$ with $I_c$ = 200 µA (Figure 3a). In the subsequent voltage sweep with $I_c$ = 300 µA, we induced an inter-QC-state switching from 2.5 $G_0$ to ~3 $G_0$ state (Figure 3b). Here, during the second sweep, the starting QC-state was found to be at 0.5 $G_0$ instead of 2.5 $G_0$. This change in state can be understood as instability of states below 3 $G_0$, as discussed above. Further, the QC-state was successively switched from 3 $G_0$ to 3.5 $G_0$ (Figure 3c), 3.5 $G_0$ to 4 $G_0$ (Figure 3d) and 4 $G_0$ to 4.5 $G_0$ (Figure 3e) during voltage sweeps with $I_c$ = 400, 500 and 600 µA, respectively. All QC-states of the device were found to be stable for at least up to 100 s (Figure 3h-j). The QC-states achieved during the inter-QC-state switching with $I_c$ of 200-500 µA either matched with the peak values from the histogram of Figure 2b-e, or fall within full-width half maxima of the peak distribution. The corresponding *I-V* traces of the conductance-voltage (*G-V*) traces shown in Figure 3a-e are shown in the supplementary Figure S3.

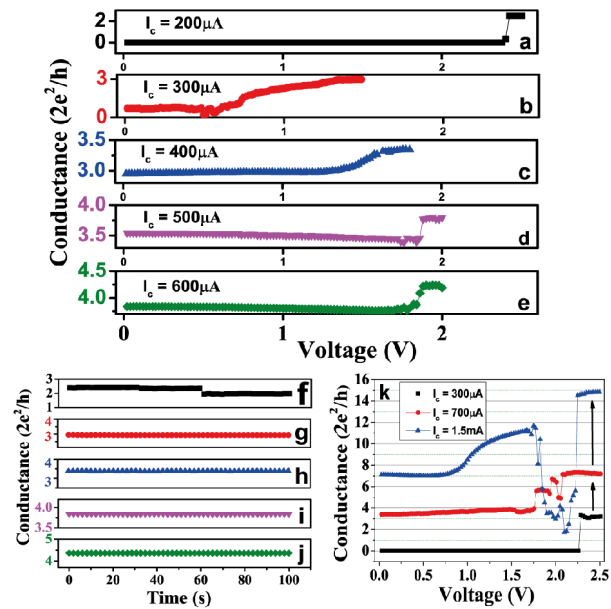

**Figure 3.** The interstate transitions between two quantized states with successive voltage sweeps during SET is exhibited in figures (a)-(e) with their corresponding final state stabilities in (f)-(j). (a) The device was first SET with $I_c$ = 200 µA, reached to 2.5 $G_0$. (b) In next voltage sweep with $I_c$ = 300 µA, a state of ~3 $G_0$ is achieved. (c) In subsequent voltage sweep with $I_c$ = 400 µA, state of ~3 $G_0$ switched to ~ 3.5 $G_0$. (d) Further increasing $I_c$ to 500 µA, switching from ~3.5 $G_0$ to ~ 4 $G_0$ is induced. (e) And subsequently, in voltage sweep with $I_c$ = 600 µA, state of 4 $G_0$ switched to 4.5 $G_0$. (f)-(j) shows the stability of QC-states reached during switching steps (a)-(e). (k) The inter-QC-state switching between two quantized states with successive voltage sweeps with $I_c$ = 300 µA, 700 µA and 1.5 mA is exhibited.

The inter-QC-state switching where the $I_c$ values were increased in steps of 400 µA and 800 µA were also performed. In Figure 3k, the device was switched to ~3 $G_0$ state with $I_c$ = 300 µA (black trace) and then in subsequent voltage sweep with $I_c$ = 700 µA, the device switched to ~7 $G_0$ state (red trace). Further, as another voltage sweep was performed with $I_c$ =

1.5 mA, the device switched from 7 $G_0$ to 15 $G_0$. While switching from 3 $G_0$ to 7 $G_0$, the device showed indications to stop at different intermediate QC-states, however, due to higher $I_c$ limit, the devices stopped only at 7 $G_0$. It appears that an $I_c$ of more than 300 µA and less than 700 µA would have possibly stabilized the device at some intermediate QC-state. During the voltage sweep with $I_c$ = 1.5 mA, the device exhibited instability around 12 $G_0$ state (Figure 3k, blue trace). Since, the device can switch in both unipolar and bipolar modes, it can be understood as the device's tendency to RESET in unipolar mode due to very high currents, however, the voltage was in the range of SET (1.5-2.5 V), thus the device switched to 15 $G_0$.

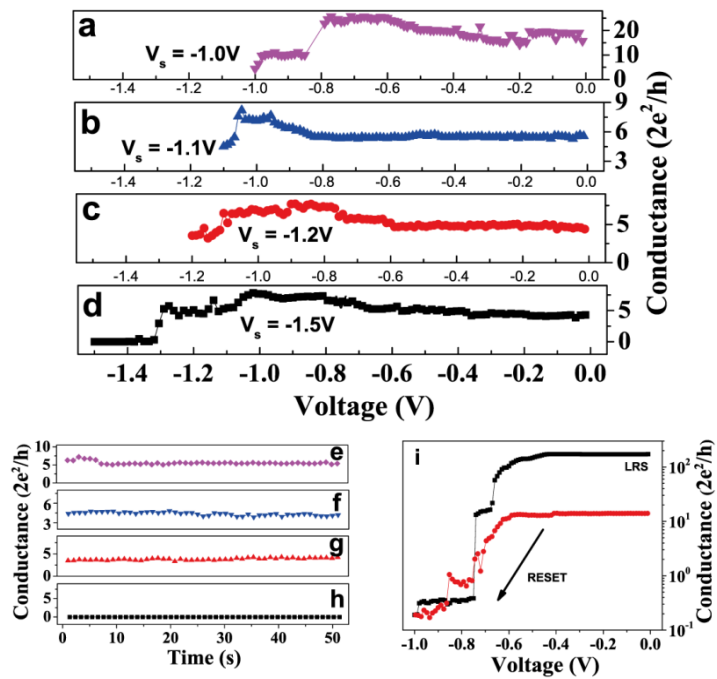

**Figure 4.** The *G-V* traces of inter-QC-state switching with successive voltage sweeps during RESET by varying the stop voltages. (a) The device in LRS is switched to ~6 $G_0$ with stop voltage of $V_s$ = -1.0 V. (b) With of $V_s$ = -1.1 V, a inter-QC-state switching from ~6 $G_0$ to ~4.5 $G_0$ is achieved. (c) Subsequently, during voltage sweep with $V_s$ = -1.2 V, QC-state switched to ~3.5 $G_0$. (d) The Figure shows a complete RESET to HRS with $V_s$ = -1.5 V. (e)-(h) The stability with time for final QC-states achieved in figures (a)-(d) are respectively shown. (i) Complete RESET transitions are exhibited. The black trace shows the RESET from LRS to HRS and the red trace shows RESET from ~15 $G_0$ to HRS state. The complete SET-RESET cycles of traces in 4i is given in the supplementary figure S5.

The inter-QC-state switching was also controlled and reproducibly performed in RESET direction. Figure 4 shows three successive steps of inter-QC-state switching of a device, where different stop voltages are used to control switching to different QC levels. The device was, firstly, SET to ~20 $G_0$ state. Then, an inter-QC-state switching from 20 $G_0$ to 6 $G_0$ was induced by a voltage sweep, where -1.0 V was kept as the stop-voltage ($V_s$), shown in Figure 4a. In the subsequent sweeps, the QC-state switched from 6 $G_0$ to 4.5 $G_0$ (Figure 4b) and from 4.5 $G_0$ to 3.5 $G_0$ (Figure 4c), with $V_s$ = -1.1 V and -1.2 V, respectively. In another subsequent sweep, the QC-state switched from 3.5 $G_0$ to a very high resistance state (i.e. complete RESET) with $V_s$ = –1.5 V, as shown in Figure 4d. Each QC-state, after every switching, was found to be stable with time (Figure 4e-h).

During RESET switching, the critical parameter was the stop-voltage instead of the current compliance limit. For example, during the voltage sweep in Figure 4a, the conductance starts to decrease or in other words, resistance of the device starts to increase at >-0.8 V. This voltage of -0.8 V becomes important, as, for any stop-voltage chosen little more than -0.8, the device stops at an intermediate stable QC-state, as shown in Figure 4a-c. However, if stop-voltage is kept sufficiently high, i.e. close to higher end of the RESET voltage range (>-1.0 V), the device will RESET completely, as shown in Figure 4i. However, this stop-voltage is not a fixed value, as devices have run-to run variations and have a range of voltage for RESET, as it is -0.4 V to -1.2 V for our devices. So, if a device starts to RESET at lower voltage (example: -0.6 V as shown in Figure 4i), and the stop-voltage is chosen to be -1.0 V, the device RESETs completely earlier than -1.0 V (Figure 4i), and thus, the device cannot be stopped at any intermediate QC-states. However, if stop-voltage would have been kept in the range -0.7 to -0.8 V for the two RESET traces in figure 4i, then the device could, possibly, have stopped at an intermediate QC-state.

Hundreds of inter-QC-state switching events in both SET and RESET directions were performed. The SET and RESET inter-QC-state switching (Figure 3 and 4) are distinguished by the limiting conditions of current compliance and stop-voltage during voltage sweep cycles, respectively. However, to ensure complete RESET from any QC-state, both current compliance as well as stop-voltage needs to be kept high.

In summary, stable and reproducible QC- states were achieved in Al/$Nb_2O_5$/Pt devices by limiting current compliance during the current-voltage measurements. All the states were stable at least for 500 s, and the higher conductance states exhibited longer retention times. The stable quantized states could be controllably switched to higher $G_0$ (SET direction) or to lower $G_0$ (RESET direction) states, by imposing the current compliance or stop-voltage limits, respectively. The conditions for complete RESET, starting from any quantized state, could also be selectively induced by lifting limiting conditions on current or voltage during RESET voltage sweep. The possibility of utilizing the QC-states in the resistive devices for multilevel logic shows potentials for achieving high-density storage.

## ASSOCIATED CONTENT

**Supporting Information**

Additional details on device fabrication, switching characteristics and stability; the inter-quantized-state switching in SET as well as in RESET directions.

## AUTHOR INFORMATION

**Corresponding author**

*E-mail: kumarajeet@nplindia.org

**Orchid**

Ajeet Kumar: 0000-0002-4878-505X

**Notes:** The authors declare no competing financial interest.

## ACKNOWLEDGEMENTS

We would like to thank Dr. Karthik Krishnan for helpful discussion. S. D. would like to thank University Grant Commission (UGC) under senior research fellowship (SRF) for financial support. This work was financially supported partly by DST INSPIRE Fellowship and partly by CSIR network project AQuaRIUS (PSC0110).## REFERENCES

(1) Pirovano, A.; Schuegraf, K. Memory Grows Up. *Nat. Nanotechnol.* **2010**, *5*, 177.

(2) Lankhorst, M. H.; Ketelaars, B. W.; Wolters, R. A. Low-Cost and Nanoscale Non-Volatile Memory Concept for Future Silicon Chips. *Nat. Mater.* **2005**, *4*, 347.

(3) Waser, R.; Aono, M. Nanoionics-Based Resistive Switching Memories. *Nat. Mater.* **2007**, *6*, 833-840.

(4) Waser, R.; Dittmann, R.; Staikov, G.; Szot, K. Redox Based Resistive Switching Memories–Nanoionic Mechanisms, Prospects, and Challenges. *Adv. Mater.* **2009**, *21*, 2632-2663.

(5) Li, Y.; Long, S.; Liu, Y.; Hu, C.; Teng, J.; Liu, Q.; Lv, H.; Suñé, J.; Liu, M. Conductance Quantization in Resistive Random Access Memory. *Nanoscale Res. Lett.* **2015**, *10*, 420.

(6) Lv, F.; Ling, K.; Wang, W.; Chen, P.; Liu, F.; Kong, W.; Zhu, C.; Liu, J.; Long, L. Multilevel Resistance Switching Behavior in $PbTiO_3$/Nb:$SrTiO_3$(100) Heterostructure Films Grown by Hydrothermal Epitaxy. *J. Alloys Compd.* **2019**, *778*, 768-773.

(7) Wu, L.; Guo, J.; Zhong, W.; Zhang, W.; Kang, X.; Chen, W.; Du, Y. Flexible, Multilevel, and Low-Operating-Voltage Resistive Memory Based on $MoS_2$–rGO Hybrid. *Appl. Surf. Sci.* **2019**, *463*, 947-952.

(8) Xu, J.; Zhao, X.; Wang, Z.; Xu, H.; Hu, J.; Ma, J.; Liu, Y. Biodegradable Natural Pectin-Based Flexible Multilevel Resistive Switching Memory for Transient Electronics. *Small* **2019**, *15*, 1803970.

(9) Vishwanath, S. K.; Woo, H.; Jeon, S. Enhancement of Resistive Switching Properties in $Al_2O_3$ Bilayer-Based Atomic Switches: Multilevel Resistive Switching. *Nanotechnology* **2018**, *29*, 235202.

(10) Kim, S.-T.; Cho, W.-J. Improvement of Multi-Level Resistive Switching Characteristics in Solution-Processed $AlO_X$-Based Non-Volatile Resistive Memory Using Microwave Irradiation. *Semicond. Sci. Technol.* **2017**, *33*, 015009.

(11) Nagareddy, V. K.; Barnes, M. D.; Zipoli, F.; Lai, K. T.; Alexeev, A. M.; Craciun, M. F.; Wright, C. D. Multilevel Ultrafast Flexible Nanoscale Nonvolatile Hybrid Graphene Oxide–Titanium Oxide Memories. *ACS Nano* **2017**, *11*, 3010-3021.

(12) Zhao, X.; Fan, Z.; Xu, H.; Wang, Z.; Xu, J.; Ma, J.; Liu, Y. Reversible Alternation between Bipolar and Unipolar Resistive Switching in Ag/$MoS_2$/Au Structure for Multilevel Flexible Memory. *J. Mater. Chem. C* **2018**, *6*, 7195-7200.

(13) Hou, P.; Gao, Z.; Ni, K. Multilevel Data Storage Memory Based on Polycrystalline $SrTiO_3$ Ultrathin Film. *RSC Adv.* **2017**, *7*, 49753-49758.

(14) Tan, T.; Du, Y.; Cao, A.; Sun, Y.; Zhang, H.; Zha, G. Resistive Switching of the $HfO_x$/$HfO_2$ Bilayer Heterostructure and Its Transmission Characteristics as a Synapse. *RSC Adv.* **2018**, *8*, 41884-41891.

(15) Xu, Z.-X.; Yan, J.-M.; Xu, M.; Guo, L.; Chen, T.-W.; Gao, G.-Y.; Wang, Y.; Li, X.-G.; Luo, H.-S.; Zheng, R.-K. Electric-Field-Controllable Nonvolatile Multilevel Resistance Switching of $Bi_{0.93}Sb_{0.07}$/PMN-0.29PT(111) Heterostructures. *Appl. Phys. Lett.* **2018**, *113*, 223504.
11

# Controlled inter-state switching between quantized conductance states in resistive devices for multilevel memory

Sweety Deswal,[†,‡] Rupali R. Malode,[§] Ashok Kumar,[†,‡] and Ajeet Kumar*[,†,‡]

[†]Academy of Scientific and Innovative Research (AcSIR), Ghaziabad- 201002, India.

[‡]CSIR-National Physical Laboratory, Dr. K.S. Krishnan Marg, New Delhi 110012, India.

[§]Maulana Azad National Institute of Technology, Bhopal, Madhya Pradesh 462003, India.

*Email: kumarajeet@nplindia.org

**Methods**

The Al/$Nb_2O_5$/Pt based resistive switching devices were fabricated by depositing ~30 nm $Nb_2O_5$ thin films over Platinised Si/$SiO_2$ substrate and subsequently depositing Al as top electrode. After standard cleaning of the substrate, the bottom electrode was masked with NiCr alloy strip of width ~0.7 mm to make Pt available for bottom electrode contact. Then, $Nb_2O_5$ thin films were grown using reactive dc magnetron sputtering at a pressure of $2.0 \times 10^{-2}$ mbar in gas mixture of Ar (94%) and $O_2$ (6%), while the base pressure was evacuated at ~$6 \times 10^{-7}$ mbar. Thereafter, the top electrode of Al was deposited with areas ranging from 100 to 500 μm$^2$ using a shadow mask by thermal evaporation technique. The current-voltage (I-V) characteristics were measured on two probe station with Agilent 2450 source-measure unit. All the measurements were performed at room temperature by applying the voltage source on the Al top electrode with the Pt bottom electrode grounded. The voltage was swept keeping a current compliance for all electrical measurements. This method of fabricating device and electrical measurement was also followed in our previous report.[s1]

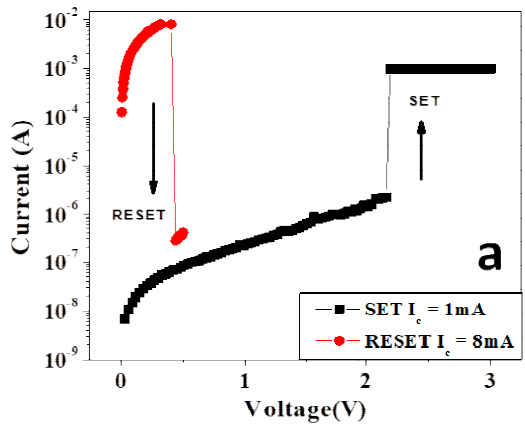

Fig. S1 The Al/Nb$_2$O$_5$/Pt device shows both unipolar (a) and bipolar (b) switching characteristics. Both (a) and (b) are consecutive switching cycles of one device. Once the electroformation is done, device can switch in both modes. After the SET voltage step with I$_c$ = 1 mA, the RESET step can be taken in either polarity of the voltage. The current required for the bipolar switching (10 mA) is similar to unipolar switching (8 mA). This is similar to previous reported observations of resistive devices showing both unipolar and bipolar switching characteristics.[s2]

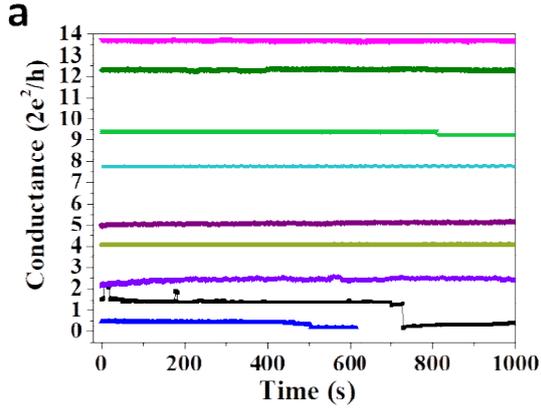

**Fig. S2 (a)** The retention time of various quantized states observed during SET, is exhibited. The states below 3 $G_0$ showed stablility for ~ 500 to 750 seconds. The weaker filament formed with smaller current compliance, results in lower stability of the states. The states ≥ 3 $G_0$ showed stability for at-least 1000 seconds. This higher stability is due to formation of stronger filament with larger current compliance values. **(b)** shows resistance of a QC-state of 1.5 $G_0$ with four different read voltages. The 1.5 $G_0$ QC-state was stable for longer time at low read voltages, up to 0.3 V, and became unstable at 0.4 V.

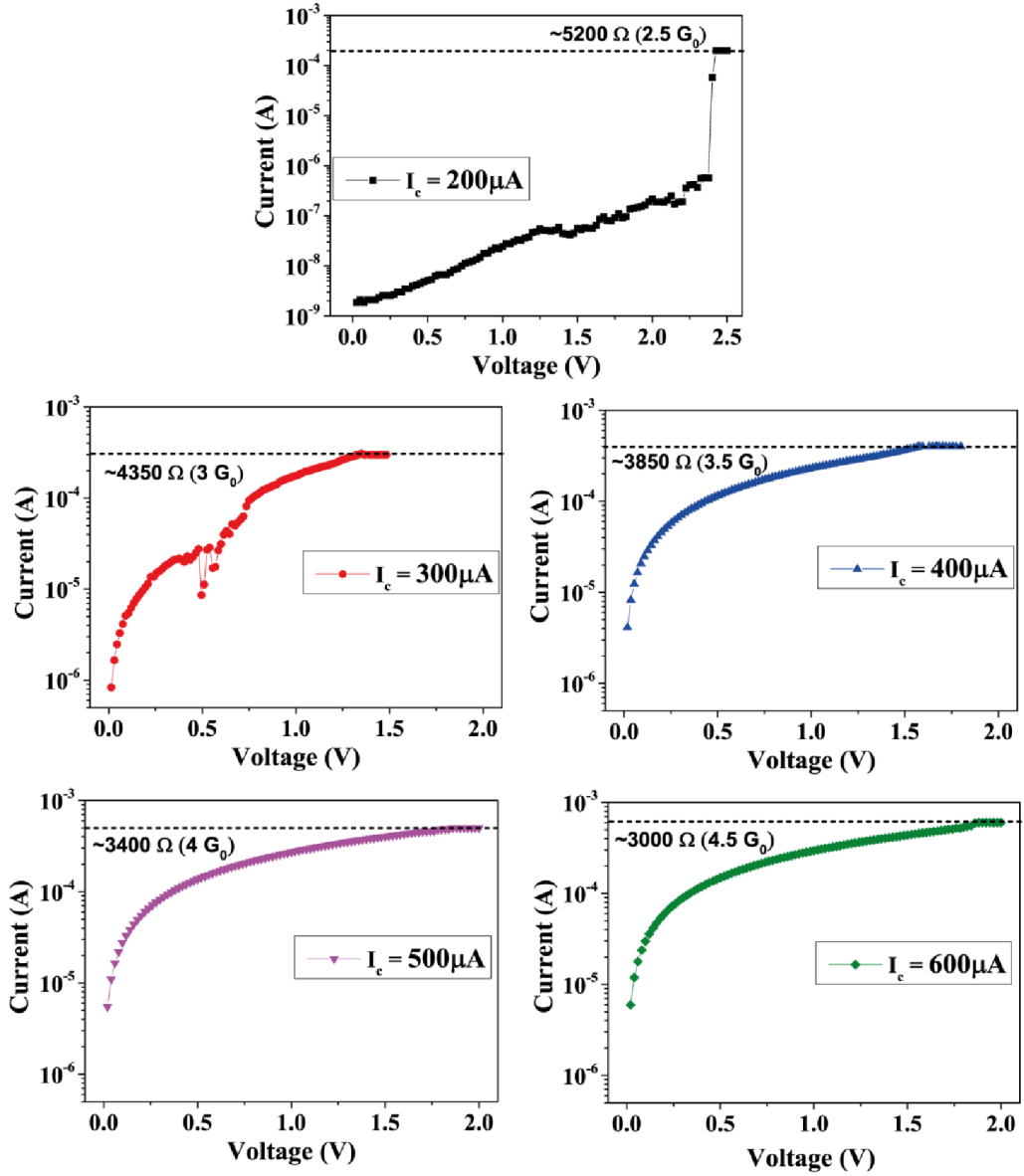

**Fig. S3** The *I-V* traces of inter-quantized state switching in the SET direction with I$_c$ = 200, 300, 400, 500, and 600 µA is shown in (a)-(e), respectively. The corresponding *G-V* traces of these curves are presented in the manuscript in Fig. 3a-e.

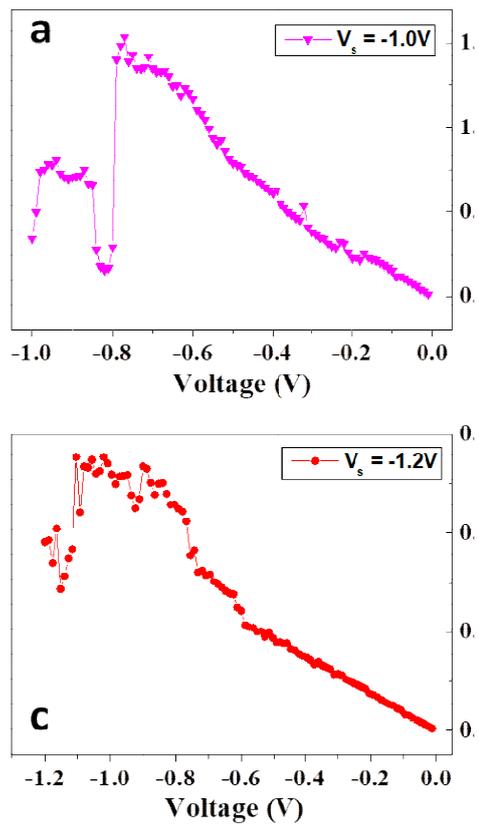

**Fig. S4** *I-V* traces of interstate switchings between quantized states with successive voltage sweeps during RESET are exhibited by varying the magnitude of stop voltages. (a)-(d) show four RESET traces with $V_s$ = -1.0, -1.1, -1.2, and -1.5 V. The corresponding *G-V* traces are presented in the manuscript in the Fig. 4a-d.

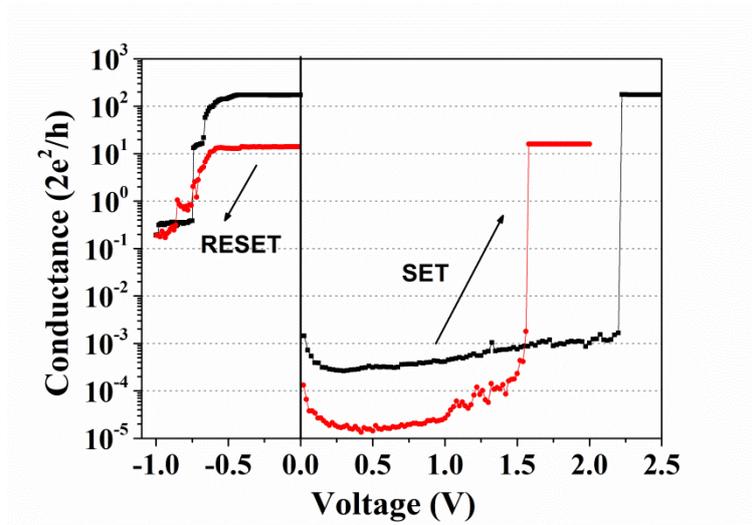

**Fig. S5** Complete SET-RESET cycles of the traces presented in the Fig. 4i.

**References**

S1  S. Deswal, A. Kumar and A. Kumar, *AIP Adv.*, 2018, **8**, 085014.

S2  Y. Sharma, S. P. Pavunny, E. Fachini, J. F. Scott, R. S. Katiyar, *J. Appl. phys.* 2015, **118**, 094506.